\documentclass{llncs}
\usepackage{version}
\usepackage{sha}

\title{Instruction Sequence Expressions for \\ the Secure Hash Algorithm
       SHA-256}
\author{J.A. Bergstra \and C.A. Middelburg}
\institute{Informatics Institute, Faculty of Science, University of
           Amsterdam, \\
           Science Park~904, 1098~XH Amsterdam, the Netherlands \\
           \email{J.A.Bergstra@uva.nl,C.A.Middelburg@uva.nl}}

\begin{document}
\maketitle

\begin{abstract}
\sloppy
The secure hash function SHA-256 is a function on bit strings.
This means that its restriction to the bit strings of any given length
can be computed by a finite instruction sequence that contains only
instructions to set and get the content of Boolean registers, forward
jump instructions, and a termination instruction.
We describe such instruction sequences for the restrictions to bit
strings of the different possible lengths by means of uniform terms from
an algebraic theory.
\begin{classcode}
E.3, F.1.1.
\end{classcode}
\end{abstract}

\section{Introduction}
\label{sect-intro}

SHA-256 is one of the hash functions defined in the Secure Hash Standard
of the U.S.\ National Institute of Standards and
Technology~\cite{NIST12a}.
To phrase it more precisely, the standard describes an algorithm that
computes the hash function SHA-256 by means of pseudo-code.
In this paper, unlike the standard, an algorithm that computes a
function is distinguished from the computed function.
SHA-256 is called a secure hash function because it is a hash function
for which it is expected to be computationally infeasible to find an
input with a given hash value and to find two different inputs with the
same hash value.
SHA-256 is implemented in some widely used security applications and
protocols, including Bitcoin~\cite{Nak08a}, S/MIME~\cite{RT10a},
TLS~\cite{DR08a}, SSH~\cite{YL06a}, and IPsec~\cite{KS05a}.

To our knowledge, the starting point of studies of the security of
SHA-256 keeps being the above-mentioned pseudo-code description of an
algorithm that computes it
(see e.g.~\cite{GH04a,MNS11a,MNS13a,MPRR06a,NB08a,SS08a}).
SHA-256 restricted to the bit strings of a given length can be computed
by a finite single-pass instruction sequence that contains only
instructions to set and get the content of Boolean registers, forward
jump instructions, and a termination instruction (see~\cite{BM13a}).
In this paper, we describe such instruction sequences for the
restrictions to bit strings of the different possible lengths by means
of uniform terms from an algebraic theory of single-pass instruction
sequences.
Thus, we provide a mathematically precise alternative to the pseudo-code
description from the standard.

In computer science, the meaning of programs usually plays a prominent
part in the explanation of many issues concerning programs.
Moreover, what is taken for the meaning of programs is mathematical by
nature.
Yet, it is customary that practitioners do not fall back on the
mathematical meaning of programs in case explanation of issues
concerning programs is needed.
They phrase their explanations from an empirical perspective.
An attempt to approach the semantics of programming languages from the
emperical perspective that a program is in essence an instruction
sequence is made in~\cite{BL02a}.
The groundwork for the approach is an algebraic theory of single-pass
instruction sequences, called program algebra, and an algebraic theory
of mathematical objects that represent the behaviours produced by
instruction sequences under execution, called basic thread algebra.

The work on an approach to programming language semantics referred to 
above initiated a line of research in which issues relating to 
various subjects from computer science are rigorously investigated 
thinking in terms of instruction sequences.
An enumeration of most papers belonging to this line of research is 
available at~\cite{SiteIS}.
The work on computational complexity presented in~\cite{BM13a,BM14e} and 
the work on algorithmic equivalence of programs presented 
in~\cite{BM14a} were prompted by the fact that, for each function on bit 
strings, its restriction to bit strings of any given length can be 
computed by a finite instruction sequence that contains only 
instructions to set and get the content of Boolean registers, forward 
jump instructions, and a termination instruction.

This fact also incited us to look for finite instruction sequences 
containing only the above-mentioned instructions that compute the 
restrictions of a well-known function on bit strings, namely SHA-256, 
to bit strings of a fixed length.
The general aim of the line of research mentioned above is to bring 
instruction sequences as a theme in computer science better into the 
picture.
This is also the general aim of the work presented in this paper.
However, different from usual in the work referred to above, the accent 
is this time on a practical problem, namely the problem to devise 
instruction sequences that compute the restrictions of SHA-256 to bit 
strings of the different possible lengths.
As in the work referred to above, this work is carried out in the
setting of program algebra.

This paper is organized as follows.
First, we survey program algebra and the particular fragment and
instantiation of it that is used in this paper (Section~\ref{sect-PGA}).
Next, we describe how we deal with $32$-bit words by means of Boolean
registers (Section~\ref{sect-words}) and how we compute the basic and
derived operations on $32$-bit words that are used in the standard to
define SHA-256 (Section~\ref{sect-opns-words}).
Then, we give the description of instruction sequences that define
SHA-256 (Section~\ref{sect-SHA-256}).
Finally, we make some concluding remarks (Section~\ref{sect-concl}).

\section{Program Algebra}
\label{sect-PGA}

In this section, we present a brief outline of \PGA\ (ProGram Algebra)
and the particular fragment and instantiation of it that is used in
the remainder of this paper.
A mathematically precise treatment can be found in~\cite{BM13a}.

The starting-point of \PGA\ is the simple and appealing perception
of a sequential program as a single-pass instruction sequence, i.e.\ a
finite or infinite sequence of instructions of which each instruction is
executed at most once and can be dropped after it has been executed or
jumped over.

It is assumed that a fixed but arbitrary set $\BInstr$ of
\emph{basic instructions} has been given.
The intuition is that the execution of a basic instruction may modify a
state and produces a reply at its completion.
The possible replies are $\False$ and $\True$.
The actual reply is generally state-dependent.
Therefore, successive executions of the same basic instruction may
produce different replies.
The set $\BInstr$ is the basis for the set of instructions that may
occur in the instruction sequences considered in \PGA.
The elements of the latter set are called \emph{primitive instructions}.
There are five kinds of primitive instructions, which are listed below:
\begin{itemize}
\item
for each $a \in \BInstr$, a \emph{plain basic instruction} $a$;
\item
for each $a \in \BInstr$, a \emph{positive test instruction} $\ptst{a}$;
\item
for each $a \in \BInstr$, a \emph{negative test instruction} $\ntst{a}$;
\item
for each $l \in \Nat$, a \emph{forward jump instruction} $\fjmp{l}$;
\item
a \emph{termination instruction} $\halt$.
\end{itemize}
We write $\PInstr$ for the set of all primitive instructions.

On execution of an instruction sequence, these primitive instructions
have the following effects:
\begin{itemize}
\item
the effect of a positive test instruction $\ptst{a}$ is that basic
instruction $a$ is executed and execution proceeds with the next
primitive instruction if $\True$ is produced and otherwise the next
primitive instruction is skipped and execution proceeds with the
primitive instruction following the skipped one --- if there is no
primitive instruction to proceed with,
inaction occurs;
\item
the effect of a negative test instruction $\ntst{a}$ is the same as
the effect of $\ptst{a}$, but with the role of the value produced
reversed;
\item
the effect of a plain basic instruction $a$ is the same as the effect
of $\ptst{a}$, but execution always proceeds as if $\True$ is produced;
\item
the effect of a forward jump instruction $\fjmp{l}$ is that execution
proceeds with the $l$th next primitive instruction of the instruction
sequence concerned --- if $l$ equals $0$ or there is no primitive
instruction to proceed with, inaction occurs;
\item
the effect of the termination instruction $\halt$ is that execution
terminates.
\end{itemize}

To build terms, \PGA\ has a constant for each primitive instruction and
two operators.
These operators are: the binary concatenation operator ${} \conc {}$ and
the unary repetition operator ${}\rep$.
We use the notation $\Conc{i = 0}{n} P_i$, where $P_0,\ldots,P_n$ are
\PGA\ terms, for the PGA term $P_0 \conc \ldots \conc P_n$.

The instruction sequences that concern us in the remainder of this paper
are the finite ones, i.e.\ the ones that can be denoted by closed \PGA\
terms in which the repetition operator does not occur.
Moreover, the basic instructions that concern us are instructions to set
and get the content of Boolean registers.
More precisely, we take the set
\pagebreak[2]
\begin{ldispl}
\set{\inbr{i}.\getbr \where i \in \Natpos} \union
\set{\outbr{i}.\setbr{b} \where i \in \Natpos \Land b \in \Bool}
\\ \;\; {} \union
\set{\auxbr{i}.\getbr \where i \in \Natpos} \union
\set{\auxbr{i}.\setbr{b} \where i \in \Natpos \Land b \in \Bool}
\end{ldispl}%
as the set $\BInstr$ of basic instructions.

Each basic instruction consists of two parts separated by a dot.
The part on the left-hand side of the dot plays the role of the name of
a Boolean register and the part on the right-hand side of the dot plays
the role of a command to be carried out on the named Boolean register.
For each $i \in \Natpos$:
\begin{itemize}
\item
$\inbr{i}$ serves as the name of the Boolean register that is used as
$i$th input register in instruction sequences;
\item
$\outbr{i}$ serves as the name of the Boolean register that is used as
$i$th output register in instruction sequences;
\item
$\auxbr{i}$ serves as the name of the Boolean register that is used as
$i$th auxiliary register in instruction sequences.
\end{itemize}
On execution of a basic instruction, the commands have the following
effects:
\begin{itemize}
\item
the effect of $\getbr$ is that nothing changes and the reply is the
content of the named Boolean register;
\item
the effect of $\setbr{\False}$ is that the content of the named Boolean
register becomes $\False$ and the reply is $\False$;
\item
the effect of $\setbr{\True}$ is that the content of the named Boolean
register becomes $\True$ and the reply is $\True$.
\end{itemize}

Let $n,m \in \Nat$, let $\funct{f}{\set{0,1}^n}{\set{0,1}^m}$, and let
$X$ be a finite instruction sequence that can be denoted by a closed
\PGA\ term in the case that $\BInstr$ is taken as specified above.
Then $X$ \emph{computes} $f$ if there exists a $k \in \Nat$ such that
for all $b_1,\ldots,b_n \in \Bool$: if $X$ is executed in an environment
with $n$ input registers, $m$ output registers, and $k$ auxiliary
registers, the content of the input registers with names
$\inbr{1},\ldots,\inbr{n}$ are $b_1,\ldots,b_n$ when execution starts,
and the content of the output registers with names
$\outbr{1},\ldots,\outbr{m}$ are $b'_1,\ldots,b'_m$ when execution
terminates, then $f(b_1,\ldots,b_n) = b'_1,\ldots,b'_m$.

\section{Dealing with $32$-Bit Words}
\label{sect-words}

This section is concerned with dealing with bit strings of length $32$
by means of Boolean registers.
It contains definitions which facilitate the description of instruction
sequences that define SHA-256 in Section~\ref{sect-SHA-256}.
In the sequel, bit strings of length $32$ will mostly be called
\emph{$32$-bit words} or shortly \emph{words}.

Let $\kappa \in \set{\mathsf{in},\mathsf{out},\mathsf{aux}}$,
let $i \in \Natpos$, and let $\kappa{:}i$ be the name of a Boolean
register.
Then $\kappa$ and $i$ are called the \emph{kind} and \emph{number} of
the Boolean register.
Successive Boolean registers are Boolean registers of the same kind
with successive numbers.
Words are stored by means of Boolean registers such that the successive
bits of a stored word are the content of successive Boolean registers
and the first bit of the word is the content of a Boolean register
whose number is in the set $\set{n \in \Nat \where n \bmod 32 = 1}$.
If a word is taken as the binary representation of a natural number, 
then the least significant bit is the first bit of the word.
\pagebreak[2]

The words that form a part of the message to which SHA-256 is to be
applied are stored in advance of the computation in input registers,
starting with the input register with number $1$, the words that form a
part of the message digest that results from applying SHA-256 are stored
during the computation in output registers, starting with the output
register with number $1$, and the words that form a part of intermediate
results that arise during the computation, such as message schedules and
hash values are stored in auxiliary registers.

It is convenient to have available the names used in the standard for 
the words of the message blocks ($M^{(i)}_j$), the message schedule 
($W_j$), the hash value ($H_j$), the working values ($a,\ldots,h$), and 
the temporary values ($T_1,T_2$) in the current setting for the Boolean 
registers that contain the first bit of these words.
It is also convenient to have available the names $D_0,\ldots,D_7$ for
the Boolean registers that contain the first bit of the words of the 
message digest, the names $t_1,\ldots,t_6,t'_1,\ldots,t'_4$ for the 
Boolean registers that contain the first bit of the words of additional 
intermediate values that are temporarily stored,%
\footnote
{The Boolean registers with names $t'_1,\ldots,t'_4$ are reserved for
 the first bit of intermediate values that arise when computing one of
 the derived operations on bit strings introduced in
 Section~\ref{sect-opns-words}.}
and the name $\nm{cb}$ for the Boolean register that contains the carry
bit that is repeatedly stored when computing the addition operation.
Therefore, we define:
\begin{ldispl}
\begin{asceqns}
M^{(i)}_j & \deq & \inbr{k}
          & \mathrm{where}\; k = 512 \mul (i - 1) + 32 \mul j + 1
 & (1 \leq i \leq 2^{55}, 0 \leq j \leq 15), \\
W_j & \deq & \auxbr{k} & \mathrm{where}\; k = 32 \mul j + 1
 & (0 \leq j \leq 63), \\
H_j & \deq & \auxbr{k} & \mathrm{where}\; k = 32 \mul j + 2049
 & (0 \leq j \leq 7), \\
\multicolumn{5}{@{}l@{}}
 {a \deq \auxbr{2305},\hfill b \deq \auxbr{2337},\hfill\!
  c \deq \auxbr{2369},\hfill d \deq \auxbr{2401},\hfill
  e \deq \auxbr{2433},} \\
\multicolumn{5}{@{}l@{}}
 {f \deq \auxbr{2465},\hfill\; g \deq \auxbr{2497},\hfill\;
  h \deq \auxbr{2529},\hfill T_1 \deq \auxbr{2561},\hfill
  T_2 \deq \auxbr{2593},} \\
\multicolumn{5}{@{}l@{}}
 {t_1 \deq \auxbr{2625},\hfill t_2 \deq \auxbr{2657},\hfill
  t_3 \deq \auxbr{2689},\hfill t_4 \deq \auxbr{2721},\hfill
  t_5 \deq \auxbr{2753},} \\
\multicolumn{5}{@{}l@{}}
 {t_6 \deq \auxbr{2785},\hfill
  t'_1 \deq \auxbr{2817},\hfill t'_2 \deq \auxbr{2849},\hfill
  t'_3 \deq \auxbr{2881},\hfill t'_4 \deq \auxbr{2913},} \\
\multicolumn{5}{@{}l@{}}
 {\nm{cb} \deq \auxbr{2945},} \\
D_j & \deq & \outbr{k} & \mathrm{where}\; k = 32 \mul j + 1
 & (0 \leq j \leq 7).
\end{asceqns}
\end{ldispl}%

It is also convenient to have available the names used in the standard
for the words of the initial hash value:
\begin{ldispl}
H^{(0)}_0 \deq 01101010000010011110011001100111\;, \\
H^{(0)}_1 \deq 10111011011001111010111010000101\;, \\
H^{(0)}_2 \deq 00111100011011101111001101110010\;, \\
H^{(0)}_3 \deq 10100101010011111111010100111010\;, \\
H^{(0)}_4 \deq 01010001000011100101001001111111\;, \\
H^{(0)}_5 \deq 10011011000001010110100010001100\;, \\
H^{(0)}_6 \deq 00011111100000111101100110101011\;, \\
H^{(0)}_7 \deq 01011011111000001100110100011001\;;
\end{ldispl}%
and the names used in the standard for the ``SHA-256 constants'':
\begin{ldispl}
\begin{aeqns}
K_0    & \deq & 01000010100010100010111110011000\;, \\
K_1    & \deq & 01110001001101110100010010010001\;, \\
      & \vdots & \\
K_{63} & \deq & 11000110011100010111100011110010\;.%
\footnotemark
\end{aeqns}
\end{ldispl}%
\footnotetext{All 64 definitions have been put into an appendix.}

\section{Computing Operations on $32$-Bit Words}
\label{sect-opns-words}

This section is concerned with computing operations on bit strings of
length $32$.
It contains definitions which facilitate the description of instruction
sequences that define SHA-256 in Section~\ref{sect-SHA-256}.

The basic operations on bit strings that are relevant to SHA-256 are
bitwise negation, bitwise conjunction, bitwise exclusive disjunction,
shift right by $n$ positions, rotate right by $n$ positions, and 
addition modulo $32$ ($0 < n < 32$).
For these operations, we define parameterized instruction sequences
computing them in case the parameters are properly instantiated (see 
below): 
\begin{ldispl}
\nm{NOT}(\srcbr{k},\dstbr{l}) \deq {}
\\ \quad
\Conc{i = 0}{31}
 (\ptst{\srcbr{k{+}i}.\getbr}         \conc 
  \ptst{\dstbr{l{+}i}.\setbr{\False}} \conc
  \dstbr{l{+}i}.\setbr{\True})\;,
\eqnsep
\nm{AND}(\srcbri{1}{k_1},\srcbri{2}{k_2},\dstbr{l}) \deq {}
\\ \quad
\Conc{i = 0}{31}
 (\ntst{\srcbri{1}{k_1{+}i}.\getbr}  \conc \fjmp{3} \conc
  \ptst{\srcbri{2}{k_2{+}i}.\getbr}  \conc 
  \ntst{\dstbr{l{+}i}.\setbr{\True}} \conc
  \dstbr{l{+}i}.\setbr{\False})\;,
\eqnsep
\nm{XOR}(\srcbri{1}{k_1},\srcbri{2}{k_2},\dstbr{l}) \deq {}
\\ \quad
\Conc{i = 0}{31}
 (\ptst{\srcbri{1}{k_1{+}i}.\getbr}   \conc \fjmp{4} \conc
  \ptst{\srcbri{2}{k_2{+}i}.\getbr}   \conc 
  \fjmp{3} \conc \fjmp{3}             \conc {}
  \ntst{\srcbri{2}{k_2{+}i}.\getbr}   \conc 
\\ \quad \phantom{\Conc{i = 0}{31} (}
  \ptst{\dstbr{l{+}i}.\setbr{\False}} \conc
  \dstbr{l{+}i}.\setbr{\True})\;,
\eqnsep
\nm{SHR}^n(\srcbr{k},\dstbr{l}) \deq {}
\\ \quad
\Conc{i = 0}{31-n} 
 (\ptst{\srcbr{k{+}n{+}i}.\getbr}    \conc 
  \ntst{\dstbr{l{+}i}.\setbr{\True}} \conc
  \dstbr{l{+}i}.\setbr{\False})      \conc {}
\\[.5ex] \quad
\Conc{i = 0}{n-1} (\dstbr{l{+}32{-}n{+}i}.\setbr{\False})\;,
\eqnsep
\nm{ROTR}^n(\srcbr{k},\dstbr{l}) \deq {}
\\ \quad
\Conc{i = 0}{31-n} 
 (\ptst{\srcbr{k{+}n{+}i}.\getbr}    \conc 
  \ntst{\dstbr{l{+}i}.\setbr{\True}} \conc
  \dstbr{l{+}i}.\setbr{\False})      \conc {}
\\[.5ex] \quad
\Conc{i = 0}{n{-}1}
 (\ptst{\srcbr{k{+}i}.\getbr}                 \conc
  \ntst{\dstbr{l{+}32{-}n{+}i}.\setbr{\True}} \conc
  \dstbr{l{+}32{-}n{+}i}.\setbr{\False})\;,
\eqnsep
\nm{ADD}(\srcbri{1}{k_1},\srcbri{2}{k_2},\dstbr{l}) \deq {}
\\ \quad
\nm{cb}.\setbr{\False} \conc {}
\\ \quad
\Conc{i = 0}{31} 
 (\ptst{\srcbri{1}{k_1{+}i}.\getbr}    \conc \fjmp{8}  \conc 
  \ptst{\srcbri{2}{k_2{+}i}.\getbr}    \conc \fjmp{8}  \conc 
  \ntst{\nm{cb}.\getbr}                \conc \fjmp{14} \conc {} 
\\ \quad \phantom{\Conc{i = 0}{31} (}
  \dstbr{l{+}i}.\setbr{\True}          \conc 
  \nm{cb}.\setbr{\False}               \conc \fjmp{13} \conc 
  \ptst{\srcbri{2}{k_2{+}i}.\getbr}    \conc \fjmp{4}  \conc 
  \ptst{\nm{cb}.\getbr} \conc \fjmp{7} \conc \fjmp{7}  \conc {} 
\\ \quad \phantom{\Conc{i = 0}{31} (} 
  \ptst{\nm{cb}.\getbr}                \conc \fjmp{5}  \conc 
  \dstbr{l{+}i}.\setbr{\False}         \conc 
  \nm{cb}.\setbr{\True} \conc \fjmp{3} \conc
  \ptst{\dstbr{l{+}i}.\setbr{\False}}  \conc
  \dstbr{l{+}i}.\setbr{\True})\;,
\end{ldispl}%
where
$s,s_1,s_2$ range over $\set{\mathsf{in},\mathsf{aux}}$,
$d$ ranges over $\set{\mathsf{aux},\mathsf{out}}$, and
$k,k_1,k_2,l$ range over $\set{n \in \Nat \where n \bmod 32 = 1}$.
For each of these parameterized instruction sequences, all but the last
parameter correspond to the operands of the operation concerned and the
last parameter corresponds to the result of the operation concerned.
Except for $\nm{ROTR}^n$, these parameterized instruction sequences 
compute the intended operations for all instantiations of their
parameters. 
$\nm{ROTR}^n$ computes the intended operation provided that the 
instantiation of the first parameter differs from the instantiation of 
the last parameter.
In this paper, this condition will always be satisfied.

\begin{proposition}
\label{prop-basic-operations-correct}
Let $n \in \Nat$ be such that $0 < n < 32$.
Then the function on bit strings of length $32$ computed by 
\begin{enumerate}
\item
$\nm{NOT}(\inbr{1},\outbr{1}) \conc \halt$ is 
bitwise negation;
\item
$\nm{AND}(\inbr{1},\inbr{33},\outbr{1}) \conc \halt$ is 
bitwise conjunction;
\item
$\nm{XOR}(\inbr{1},\inbr{33},\outbr{1}) \conc \halt$ is
bitwise exclusive disjunction;
\item
$\nm{SHR}^n(\inbr{1},\outbr{1}) \conc \halt$ is
shift right by $n$ positions;
\item
$\nm{ROTR}^n(\inbr{1},\outbr{1}) \conc \halt$ is 
rotate right by $n$ positions;
\item
$\nm{ADD}(\inbr{1},\inbr{33},\outbr{1}) \conc \halt$
models addition modulo $2^{32}$ on natural numbers less than $2^{32}$ 
with respect to their binary representation by $32$-bit words.
\end{enumerate}
\end{proposition}
\begin{proof}
Except for the last property, these properties are easy to prove by
taking an arbitrary word position $j$ ($0 \leq j \leq 31$), making a 
case distinction on the contents of the input registers containing the
bits of the operands at position $j$, and using universal 
generalization.
In the case of the fourth and fifth property, a distinction between the 
cases $j \leq 31 - n$ and $j > 31 - n$ is needed too.
The last property is an instance of a more general property proved 
in~\cite{BM13c}.
\qed
\end{proof}

In the standard, for SHA-256, six derived operations on bit strings are
defined in terms of the above-mentioned basic operations.%
\footnote
{In the standard, basic operations and derived operations are called
 operations and functions, respectively.
}
For these operations, we also define parameterized instruction sequences
computing them:
\begin{ldispl}
\nm{CH}(\srcbri{1}{k_1},\srcbri{2}{k_2},\srcbri{3}{k_3},\dstbr{l})
 \deq {}
\\ \quad
\nm{NOT}(\srcbri{1}{k_1},t'_1) \conc
\nm{AND}(\srcbri{1}{k_1},\srcbri{2}{k_2},t'_2) \conc
\nm{AND}(t'_1,\srcbri{3}{k_3},t'_3) \conc {}
\\ \quad
\nm{XOR}(t'_2,t'_3,\dstbr{l})\;,
\eqnsep
\nm{MAJ}(\srcbri{1}{k_1},\srcbri{2}{k_2},\srcbri{3}{k_3},\dstbr{l})
 \deq {}
\\ \quad
\nm{AND}(\srcbri{1}{k_1},\srcbri{2}{k_2},t'_1) \conc
\nm{AND}(\srcbri{1}{k_1},\srcbri{3}{k_3},t'_2) \conc
\nm{AND}(\srcbri{2}{k_2},\srcbri{3}{k_3},t'_3) \conc {}
\\ \quad
\nm{XOR}(t'_1,t'_2,t'_4) \conc \nm{XOR}(t'_3,t'_4,\dstbr{l})\;,
\eqnsep
\Sigma_0(\srcbr{k},\dstbr{l})
 \deq {} 
\\ \quad
\nm{ROTR}^2(\srcbr{k},t'_1) \conc
\nm{ROTR}^{13}(\srcbr{k},t'_2) \conc
\nm{ROTR}^{22}(\srcbr{k},t'_3) \conc {}
\\ \quad
\nm{XOR}(t'_1,t'_2,t'_4) \conc \nm{XOR}(t'_3,t'_4,\dstbr{l})\;,
\end{ldispl}%
\begin{ldispl}
\Sigma_1(\srcbr{k},\dstbr{l})
 \deq {}  \hsp{24.4}
\\ \quad
\nm{ROTR}^6(\srcbr{k},t'_1) \conc
\nm{ROTR}^{11}(\srcbr{k},t'_2) \conc
\nm{ROTR}^{25}(\srcbr{k},t'_3) \conc {}
\\ \quad
\nm{XOR}(t'_1,t'_2,t'_4) \conc \nm{XOR}(t'_3,t'_4,\dstbr{l})\;,
\eqnsep
\sigma_0(\srcbr{k},\dstbr{l})
 \deq {}
\\ \quad
\nm{ROTR}^7(\srcbr{k},t'_1) \conc
\nm{ROTR}^{18}(\srcbr{k},t'_2) \conc
\nm{SHR}^3(\srcbr{k},t'_3) \conc {}
\\ \quad
\nm{XOR}(t'_1,t'_2,t'_4) \conc \nm{XOR}(t'_3,t'_4,\dstbr{l})\;,
\eqnsep
\sigma_1(\srcbr{k},\dstbr{l})
 \deq {}
\\ \quad
\nm{ROTR}^{17}(\srcbr{k},t'_1) \conc
\nm{ROTR}^{19}(\srcbr{k},t'_2) \conc
\nm{SHR}^{10}(\srcbr{k},t'_3) \conc {}
\\ \quad
\nm{XOR}(t'_1,t'_2,t'_4) \conc \nm{XOR}(t'_3,t'_4,\dstbr{l})\;,
\end{ldispl}%
where
$s,s_1,s_2,s_3$ range over $\set{\mathsf{in},\mathsf{aux}}$,
$d$ ranges over $\set{\mathsf{aux},\mathsf{out}}$,
$k,k_1,k_2,k_3,l$ range over $\set{n \in \Nat \where n \bmod 32 = 1}$.

\begin{proposition}
\label{prop-derived-operations-correct}
\sloppy
Let $n \in \Nat$ be such that $0 < n < 32$.
Then the functions on bit strings of length $32$ computed by 
the instruction sequences
$\nm{CH}(\inbr{1},\inbr{33},\inbr{65},\outbr{1})  \conc \halt$,
$\nm{MAJ}(\inbr{1},\inbr{33},\inbr{65},\outbr{1}) \conc \halt$,
$\Sigma_0(\inbr{1},\outbr{1}) \conc \halt$,
$\Sigma_1(\inbr{1},\outbr{1}) \conc \halt$,
$\sigma_0(\inbr{1},\outbr{1}) \conc \halt$, and
$\sigma_1(\inbr{1},\outbr{1}) \conc \halt$ 
are the functions with the same names defined 
in~\textup{\cite{NIST12a}}.
\end{proposition}
\begin{proof}
This follows immediately from 
Proposition~\ref{prop-basic-operations-correct} and the definitions 
of the functions concerned in~\cite{NIST12a}.
\qed
\end{proof}

Furthermore, SHA-256 also involves storing $32$-bit words and 
transferring stored $32$-bit words.
Therefore, we define the following parameterized instruction sequences:
\begin{ldispl}
\nm{SET}(b_0 \ldots b_{31},\dstbr{l}) \deq
\Conc{i = 0}{31} (\dstbr{l{+}i}.\setbr{b_{31-i}})\;,
\eqnsep
\nm{MOV}(\srcbr{k},\dstbr{l}) \deq
\Conc{i = 0}{31}
 (\ptst{\srcbr{k{+}i}.\getbr}        \conc 
  \ntst{\dstbr{l{+}i}.\setbr{\True}} \conc
  \dstbr{l{+}i}.\setbr{\False})\;,
\end{ldispl}%
\sloppy
where
$b_0,\ldots,b_{31}$ range over $\set{\False,\True}$,
$s$ ranges over $\set{\mathsf{in},\mathsf{aux}}$,
$d$ ranges over $\set{\mathsf{aux},\mathsf{out}}$, and
$k,l$ range over $\set{n \in \Nat \where n \bmod 32 = 1}$.

\begin{proposition}
\label{prop-transfer-operations-correct}
The function on bit strings of length $32$ computed by 
\begin{enumerate}
\item
$\nm{SET}(b_0 \ldots b_{31},\outbr{1}) \conc \halt$ is the bit string 
constant $b_0 \ldots b_{31}$;
\item
$\nm{MOV}(\inbr{1},\outbr{1}) \conc \halt$ is the identity function on 
bit strings of length $32$.
\end{enumerate}
\end{proposition}
\begin{proof}
These properties are easy to prove by taking an arbitrary word 
position~$j$ ($0 \leq j \leq 31$), making a case distinction on $b_j$ 
and the content of the input register containing the bit of the operand 
at position $j$, respectively, and using universal generalization.
\qed
\end{proof}

The calculation of the lengths of the parameterized instruction
sequences defined above is a matter of simple additions, subtractions, 
and multiplications.
The lengths of the instruction sequences corresponding to the basic
operations on bit strings relevant to SHA-256 are as follows:
\begin{ldispl}
\len(\nm{NOT}(\srcbr{k},\dstbr{l})) = 96\;, \\
\len(\nm{AND}(\srcbri{1}{k_1},\srcbri{2}{k_2},\dstbr{l})) = 160\;, \\
\len(\nm{XOR}(\srcbri{1}{k_1},\srcbri{2}{k_2},\dstbr{l})) = 256\;, \\
\len(\nm{SHR}^n(\srcbr{k},\dstbr{l})) = 96 - 2 \mul n\;, \\
\len(\nm{ROTR}^n(\srcbr{k},\dstbr{l})) = 96\;, \\
\len(\nm{ADD}(\srcbri{1}{k_1},\srcbri{2}{k_2},\dstbr{l})) = 673\;; \hsp{3}
\end{ldispl}%
the lengths of the instruction sequences corresponding to the derived
operations on bit strings defined in the standard are as follows:
\begin{ldispl}
\len(\nm{CH}(\srcbri{1}{k_1},\srcbri{2}{k_2},\srcbri{3}{k_3},\dstbr{l}))
 = 672\;, \\
\len(\nm{MAJ}(\srcbri{1}{k_1},\srcbri{2}{k_2},\srcbri{3}{k_3},\dstbr{l}))
 = 992\;, \\
\len(\Sigma_0(\srcbr{k},\dstbr{l})) = 800\;, \\
\len(\Sigma_1(\srcbr{k},\dstbr{l})) = 800\;, \\
\len(\sigma_0(\srcbr{k},\dstbr{l})) = 794\;, \\
\len(\sigma_1(\srcbr{k},\dstbr{l})) = 780\;;
\end{ldispl}%
and the lengths of the $\nm{SET}$ and $\nm{MOV}$ instruction sequences
are as follows:
\begin{ldispl}
\len(\nm{SET}(b_0 \ldots b_{31},\dstbr{l})) = 32\;, \\
\len(\nm{MOV}(\srcbr{k},\dstbr{l})) = 96\;. \hsp{6}
\end{ldispl}%

In the description of instruction sequences that define SHA-256 in 
Section~\ref{sect-SHA-256}, we will also use the abbreviation
\begin{ldispl}
\mathsf{CONC\;FOR}\; i = l \;\mathsf{TO}\; l': \{ P_i \}
\quad \mathrm{for} \quad
P_l \conc \ldots \conc P_{l'}\;,
\end{ldispl}%
where $l,l' \in \Nat$ are such that $l < l'$, and $P_l,\ldots,P_{l'}$
are instruction sequences. 
\linebreak[2]
We write $\mathsf{CONC\;FOR}$ instead of $\mathsf{FOR}$ to emphasize
that we have to do here with an abbreviation for the concatenation of
two or more instruction sequences.

\section{SHA-256 Hash Computation}
\label{sect-SHA-256}

In this section, we give the description of instruction sequences that
define SHA-256 using the definitions given in Sections~\ref{sect-words}
and~\ref{sect-opns-words}.

The padding of messages to a bit length that is a multiple of $512$ is
left out.
It is assumed that messages are already padded.
Thus, the bit length of a message is always a multiple of $512$.
Suppose that $N$ is the bit length of a message divided by $512$.
Because the maximum bit length of a message is $2^{64}$, we have that
$1 \leq N \leq 2^{55}$.

We write $\mathcal{M}_N$, where $1 \leq N \leq 2^{55}$, for
$\set{0,1}^{512 \mul N}$, and we write $\mathcal{M}$ for
$\Union \set{\mathcal{M}_N \where 1 \leq N \leq 2^{55}}$.
Moreover, we write $\mathcal{D}$ for $\set{0,1}^{256}$.
SHA-256 is a function from $\mathcal{M}$ to $\mathcal{D}$.
We write $\textrm{SHA-256}_N$ for the restriction of SHA-256 to
$\mathcal{M}_N$.
Clearly, SHA-256 is the unique function from $\mathcal{M}$ to
$\mathcal{D}$ such that, for each $N$ with $1 \leq N \leq 2^{55}$,
for each $w \in \mathcal{M}_N$,
$\textrm{SHA-256}(w) = \textrm{SHA-256}_N(w)$.

In Table~\ref{table-inseq-SHA-256}, an instruction sequence
$\mathrm{IS}_{\mathrm{SHA}\textrm{-}\mathrm{256}_N}$ is uniformly
described for all $N$ with $1 \leq N \leq 2^{55}$.
\begin{table}[p]
\caption{\normalsize The instruction sequence
  $\mathrm{IS}_{\mathrm{SHA}\textrm{-}\mathrm{256}_N}$}
\label{table-inseq-SHA-256}
\centering
\normalsize 
\renewcommand{\arraystretch}{1.3}
$
\begin{array}{@{}l@{}}
\mathsf{CONC\;FOR}\; j = 0 \;\mathsf{TO}\; 7:
\\
\quad\{
  \\ \phantom{\quad\{}
  \nm{SET}(H^{(0)}_i,H_i)
  \\
\quad\} \conc {}
\\
\mathsf{CONC\;FOR}\; i = 1 \;\mathsf{TO}\; N:
\\
\quad\{
  \\ \phantom{\quad\{}
  \mathsf{CONC\;FOR}\; j = 0  \;\mathsf{TO}\; 15:
  \\ \phantom{\quad\{}
  \quad\{
    \\ \phantom{\quad\{\quad\{}
    \nm{MOV}(M^{(i)}_j, W_j)
    \\ \phantom{\quad\{}
  \quad\} \conc {}
  \\ \phantom{\quad\{}
  \mathsf{CONC\;FOR}\; j = 16 \;\mathsf{TO}\; 63:
  \\ \phantom{\quad\{}
  \quad\{
    \\ \phantom{\quad\{\quad\{}
    \sigma_1(W_{j-2},t_1)  \conc \sigma_0(W_{j-15},t_2) \conc {}
    \\ \phantom{\quad\{\quad\{}
    \nm{ADD}(t_1,W_{j-7},t_3) \conc \nm{ADD}(t_2,W_{j-16},t_4) \conc
    \nm{ADD}(t_3,t_4,W_j)
    \\ \phantom{\quad\{}
  \quad\} \conc {}
  \\ \phantom{\quad\{}
  \nm{MOV}(H_0,a) \conc \nm{MOV}(H_1,b) \conc \nm{MOV}(H_2,c) \conc
  \nm{MOV}(H_3,d) \conc {}
  \\ \phantom{\quad\{}
  \nm{MOV}(H_4,e) \conc \nm{MOV}(H_5,f) \conc \nm{MOV}(H_6,g) \conc
  \nm{MOV}(H_7,h) \conc {}
  \\ \phantom{\quad\{}
  \mathsf{CONC\;FOR}\; j = 0 \;\mathsf{TO}\; 63:
  \\ \phantom{\quad\{}
  \quad\{
    \\ \phantom{\quad\{\quad\{}
    \Sigma_1(e,t_1) \conc \nm{CH}(e,f,g,t_2) \conc
    \nm{SET}(K_j,t_3) \conc {}
    \\ \phantom{\quad\{\quad\{}
    \nm{ADD}(t_1,h,t_4) \conc \nm{ADD}(t_2,t_3,t_5) \conc
    \nm{ADD}(t_5,W_j,t_6) \conc \nm{ADD}(t_4,t_6,T_1) \conc {}
    \\ \phantom{\quad\{\quad\{}
    \Sigma_0(a,t_1) \conc \nm{MAJ}(a,b,c,t_2) \conc
    \nm{ADD}(t_1,t_2,T_2) \conc {}
    \\ \phantom{\quad\{\quad\{}
    \nm{MOV}(g,h) \conc \nm{MOV}(f,g) \conc \nm{MOV}(e,f) \conc
    \nm{ADD}(d,T_1,e) \conc {}
    \\ \phantom{\quad\{\quad\{}
    \nm{MOV}(c,d) \conc \nm{MOV}(b,c) \conc \nm{MOV}(a,b) \conc
    \nm{ADD}(T_1,T_2,a)
    \\ \phantom{\quad\{}
  \quad\} \conc {}
  \\ \phantom{\quad\{}
  \nm{ADD}(a,H_0,H_0) \conc \nm{ADD}(b,H_1,H_1) \conc 
  \nm{ADD}(c,H_2,H_2) \conc \nm{ADD}(d,H_3,H_3) \conc 
  \\ \phantom{\quad\{}
  \nm{ADD}(e,H_4,H_4) \conc \nm{ADD}(f,H_5,H_5) \conc 
  \nm{ADD}(g,H_6,H_6) \conc \nm{ADD}(h,H_7,H_7)
  \\
\quad\} \conc {}
\\
\mathsf{CONC\;FOR}\; j = 0 \;\mathsf{TO}\; 7:
\\
\quad\{
  \\ \phantom{\quad\{}
  \nm{MOV}(H_j,D_j)
  \\
\quad\} \conc {}
\\
\halt
\end{array}
$
\end{table}
\begin{varclaim}
For each $N$ with $1 \leq N \leq 2^{55}$, the instruction
sequence $\mathrm{IS}_{\mathrm{SHA}\textrm{-}\mathrm{256}_N}$ computes
the function {\upshape$\textrm{SHA-256}_N$}.
\end{varclaim}
In the standard, the function SHA-256 is defined indirectly and 
informally by means of a pseudo-code description of an algorithm that 
computes it.
Therefore, we cannot formally prove this claim.
However, we follow the description of the algorithm in the standard so 
precisely in the description of 
$\mathrm{IS}_{\mathrm{SHA}\textrm{-}\mathrm{256}_N}$ 
that, given Propositions~\ref{prop-basic-operations-correct}, 
\ref{prop-derived-operations-correct}, 
and~\ref{prop-transfer-operations-correct}, 
the claim is unlikely to be wrong unless the pseudo code should not be 
interpreted as to be expected.

An easy calculation leads to the following result.
\begin{proposition}
\label{prop-SHA-length}
For each $N$ with $1 \leq N \leq 2^{55}$, the length of the
instruction sequence $\mathrm{IS}_{\mathrm{SHA}\textrm{-}\mathrm{256}_N}$
is $729464 \mul N + 1025$.
\end{proposition}
\begin{proof}
The calculation is a matter of simple additions and multiplications,
using the lengths of the parameterized instruction sequences given in
Section~\ref{sect-opns-words}:
\begin{ldispl}
 8 \mul 32 + {} 
 \\
 N \mul (16 \mul 96 +  {} 
 \\ \phantom{N \mul (}
         48 \mul (780 + 794 + 3 \mul 673) + {} 
         \\ \phantom{N \mul (}
         8 \mul 96 + {}
         \\ \phantom{N \mul (}
         64 \mul (800 + 672 + 32 + 4 \mul 673 + {}
         \\ \phantom{N \mul (64 \mul (}
                  800 + 992 + 673 + {}
                  \\ \phantom{N \mul (64 \mul (}
                  3 \mul 96 + 673 + 3 \mul 96 + 673) + {}
                  \\ \phantom{N \mul (}
          8 \mul 673) + {}
          \\
 8 \mul 96 + {}
 \\
 1 
 \\
 {} = {} 
 \\
 729464 \mul N + 1025\;.
\end{ldispl}%
The left-hand side of this equation is laid out in such a way that the
structure of the description in Table~\ref{table-inseq-SHA-256} is
clearly reflected.
\qed
\end{proof}

Recall that the instruction sequence
$\mathrm{IS}_{\mathrm{SHA}\textrm{-}\mathrm{256}_N}$
($1 \leq N \leq 2^{55}$) contains only instructions to set and get the
content of Boolean registers, forward jump instructions, and a
termination instruction.
Results from~\cite{BM13a} suggest that, in the case of instruction 
sequences of this kind, instruction sequence size and computation time 
are polynomially related complexity measures.
Notice that, if the message has the maximum bit length
($\pm 1.8 \mul {10}^{19}$), the length of the instruction sequence is
$\pm 2.6 \mul {10}^{22}$.

The maximum number of input registers needed is $2^{64}$ and the number
of output registers needed is $256$.
The number of auxiliary registers used is $2945$.
We expect that the number of auxiliary registers that are used by an 
instruction sequence and computation space are related complexity 
measures.
Notice that the number of auxiliary registers used does not depend on 
the length of the message.

\section{Concluding Remarks}
\label{sect-concl}

We have described instruction sequences that compute the restrictions of
the secure hash function SHA-256 to the bit strings of the different
possible lengths by means of uniform terms from the algebraic theory of
single-pass instruction sequences known as \PGA.
Thus, we have provided a mathematically precise alternative to the
pseudo-code description of an algorithm that computes SHA-256 found in
the standard.

In previous work belonging to the line of research initiated by the work
presented in~\cite{BL02a}, the work almost always concerns rigorous 
investigation of theoretical issues thinking in terms of instruction 
sequences (see~\cite{SiteIS}).
This may give the impression that \PGA\ is only suitable for work of 
that kind.
The use of \PGA\ in the work presented in this paper shows that it is
more versatile.
However, this work has also shown that scalability calls for extension
of \PGA\ to an instruction sequence calculus that includes among other
things a variable binding generalized concatenation operator and a
suitable definition mechanism.

Results from~\cite{BM13a} suggest that, in the case of instruction 
sequences of the kind that we have dealt with in this paper, instruction 
sequence size and computation time are polynomially related complexity 
measures.
An option for future work is investigating the possible role of this
complexity measure in issues concerning the complexity of the different
kinds of attack on secure hash functions like SHA-256.

\subsection*{Acknowledgements}
We thank Bob Diertens from the University of Amsterdam for carefully
reading an earlier version of this paper, pointing out annoying errors
in it, and developing programs by which the description of
$\mathrm{IS}_{\mathrm{SHA}\textrm{-}\mathrm{256}_N}$ given in this
paper can be transformed into an instruction sequence that can be
executed by means of the PGA toolset~\cite{Die03a}.
\pagebreak

\bibliographystyle{splncs03}
\bibliography{IS}

\begin{thebibliography}{10}
\providecommand{\url}[1]{\texttt{#1}}
\providecommand{\urlprefix}{URL }

\bibitem{BL02a}
Bergstra, J.A., Loots, M.E.: Program algebra for sequential code. Journal of
  Logic and Algebraic Programming  51(2),  125--156 (2002)

\bibitem{BM13c}
Bergstra, J.A., Middelburg, C.A.: Instruction sequence expressions for the
  {Karatsuba} multiplication algorithm. {\tt arXiv:1312.1529v2 [cs.PL]}
  (2013)

\bibitem{BM13a}
Bergstra, J.A., Middelburg, C.A.: Instruction sequence based non-uniform
  complexity classes. Scientific Annals of Computer Science  24(1),  47--89
  (2014)

\bibitem{BM14a}
Bergstra, J.A., Middelburg, C.A.: On algorithmic equivalence of instruction
  sequences for computing bit string functions. Fundamenta Informaticae
  138(4),  411--434 (2015)

\bibitem{BM14e}
Bergstra, J.A., Middelburg, C.A.: Instruction sequence size complexity of
  parity. Fundamenta Informaticae  149(3),  297--309 (2016)

\bibitem{DR08a}
Dierks, T., Rescorla, E.: {The Transport Layer Security (TLS) Protocol Version
  1.2}. The Internet Society, IETF RFC 5246 (2008)

\bibitem{Die03a}
Diertens, B.: A toolset for {PGA}. Electronic Report PRG0302, Programming
  Research Group, University of Amsterdam (2003), available at {\tt
  http://www.science.\linebreak[2]uva.nl/research/prog/publications.html}

\bibitem{GH04a}
Gilbert, H., Handschuh, H.: Security analysis of {SHA-256} and sisters. In:
  Matsui, M., Zuccherato, R. (eds.) SAC 2003. Lecture Notes in Computer
  Science, vol. 3006, pp. 175--193. Springer-Verlag (2004)

\bibitem{KS05a}
Kent, S., Seo, K.: {Security Architecture for the Internet Protocol}. The
  Internet Society, IETF RFC 4301 (2005)

\bibitem{MNS11a}
Mendel, F., Nad, T., Schl{\"{a}}ffer, M.: Finding {SHA-2} characteristics:
  Searching through a minefield of contradictions. In: Lee, D.H., Wang, X.
  (eds.) ASIACRYPT 2011. Lecture Notes in Computer Science, vol. 7073, pp.
  288--307. Springer-Verlag (2011)

\bibitem{MNS13a}
Mendel, F., Nad, T., Schl{\"{a}}ffer, M.: Improving local collisions: New
  attacks on reduced {SHA-256}. In: Johansson, T., Nguyen, P. (eds.) EUROCRYPT
  2013. Lecture Notes in Computer Science, vol. 7881, pp. 262--278.
  Springer-Verlag (2013)

\bibitem{MPRR06a}
Mendel, F., Pramstaller, N., Rechberger, C., Rijmen, V.: Analysis of
  step-reduced {SHA-256}. In: Robshaw, M.J.B. (ed.) FSE 2006. Lecture Notes in
  Computer Science, vol. 4047, pp. 126--143. Springer-Verlag (2006)

\bibitem{SiteIS}
Middelburg, C.A.: Instruction sequences as a theme in computer science. {\tt
  https:\linebreak[2]//instructionsequence.wordpress.com/} (2015)

\bibitem{Nak08a}
Nakamoto, S.: Bitcoin: A peer-to-peer electronic cash system. {\tt
  http://bitcoin.\linebreak[2]org/bitcoin.pdf} (2008)

\bibitem{NB08a}
Nikoli{\'{c}}, I., Biryukov, A.: Collisions for step-reduced {SHA-256}. In:
  Nyberg, K. (ed.) FSE 2008. Lecture Notes in Computer Science, vol. 5086, pp.
  1--15. Springer-Verlag (2008)

\bibitem{RT10a}
Ramsdell, B., Turner, S.: {Secure/Multipurpose Internet Mail Extensions
  (S/\linebreak[2]MIME) Version 3.2 Message Specification}. The Internet
  Society, IETF RFC 5751 (2010)

\bibitem{SS08a}
Sanadhya, S.K., Sarkar, P.: New collision attacks against up to 24-step
  {SHA-256}. In: Chowdhury, D.R., Rijmen, V., Das, A. (eds.) INDOCRYPT 2008.
  Lecture Notes in Computer Science, vol. 5365, pp. 91--103. Springer-Verlag
  (2008)

\bibitem{YL06a}
Ylonen, T., Lonvick, C.: {The Secure Shell (SSH) Transport Layer Protocol}. The
  Internet Society, IETF RFC 4253 (2006)

\bibitem{NIST12a}
{Secure Hash Standard}. National Institute of Standards and Technology, FIPS
  PUB 180-4 (2012)

\end{thebibliography}

\appendix

\section{Definitions of the SHA-256 constants}

\begin{ldispl}
\begin{aeqns}
K_0      & \deq & 01000010100010100010111110011000\;, \\
K_1      & \deq & 01110001001101110100010010010001\;, \\
K_2      & \deq & 10110101110000001111101111001111\;, \\
K_3      & \deq & 11101001101101011101101110100101\;, \\
K_4      & \deq & 00111001010101101100001001011011\;, \\
K_5      & \deq & 01011001111100010001000111110001\;, \\
K_6      & \deq & 10010010001111111000001010100100\;, \\
K_7      & \deq & 10101011000111000101111011010101\;, \\
K_8      & \deq & 11011000000001111010101010011000\;, \\
K_9      & \deq & 00010010100000110101101100000001\;, \\
K_{10}   & \deq & 00100100001100011000010110111110\;, \\
K_{11}   & \deq & 01010101000011000111110111000011\;, \\
K_{12}   & \deq & 01110010101111100101110101110100\;, \\
K_{13}   & \deq & 10000000110111101011000111111110\;, \\
K_{14}   & \deq & 10011011110111000000011010100111\;, \\
K_{15}   & \deq & 11000001100110111111000101110100\;, \\
K_{16}   & \deq & 11100100100110110110100111000001\;, \\
K_{17}   & \deq & 11101111101111100100011110000110\;, \\
K_{18}   & \deq & 00001111110000011001110111000110\;, \\
K_{19}   & \deq & 00100100000011001010000111001100\;, \\
K_{20}   & \deq & 00101101111010010010110001101111\;, \\
K_{21}   & \deq & 01001010011101001000010010101010\;, \\
K_{22}   & \deq & 01011100101100001010100111011100\;, \\
K_{23}   & \deq & 01110110111110011000100011011010\;, \\
K_{24}   & \deq & 10011000001111100101000101010010\;, \\
K_{25}   & \deq & 10101000001100011100011001101101\;, \\
K_{26}   & \deq & 10110000000000110010011111001000\;, \\
K_{27}   & \deq & 10111111010110010111111111000111\;, \\
K_{28}   & \deq & 11000110111000000000101111110011\;, \\
K_{29}   & \deq & 11010101101001111001000101000111\;, \\
K_{30}   & \deq & 00000110110010100110001101010001\;, \\
K_{31}   & \deq & 00010100001010010010100101100111\;, \\
K_{32}   & \deq & 00100111101101110000101010000101\;, \\
K_{33}   & \deq & 00101110000110110010000100111000\;, \\
K_{34}   & \deq & 01001101001011000110110111111100\;,
\end{aeqns}
\end{ldispl}%
\begin{ldispl}
\begin{aeqns}
K_{35}   & \deq & 01010011001110000000110100010011\;, \\
K_{36}   & \deq & 01100101000010100111001101010100\;, \\
K_{37}   & \deq & 01110110011010100000101010111011\;, \\
K_{38}   & \deq & 10000001110000101100100100101110\;, \\
K_{39}   & \deq & 10010010011100100010110010000101\;, \\
K_{40}   & \deq & 10100010101111111110100010100001\;, \\
K_{41}   & \deq & 10101000000110100110011001001011\;, \\
K_{42}   & \deq & 11000010010010111000101101110000\;, \\
K_{43}   & \deq & 11000111011011000101000110100011\;, \\
K_{44}   & \deq & 11010001100100101110100000011001\;, \\
K_{45}   & \deq & 11010110100110010000011000100100\;, \\
K_{46}   & \deq & 11110100000011100011010110000101\;, \\
K_{47}   & \deq & 00010000011010101010000001110000\;, \\
K_{48}   & \deq & 00011001101001001100000100010110\;, \\
K_{49}   & \deq & 00011110001101110110110000001000\;, \\
K_{50}   & \deq & 00100111010010000111011101001100\;, \\
K_{51}   & \deq & 00110100101100001011110010110101\;, \\
K_{52}   & \deq & 00111001000111000000110010110011\;, \\
K_{53}   & \deq & 01001110110110001010101001001010\;, \\
K_{54}   & \deq & 01011011100111001100101001001111\;, \\
K_{55}   & \deq & 01101000001011100110111111110011\;, \\
K_{56}   & \deq & 01110100100011111000001011101110\;, \\
K_{57}   & \deq & 01111000101001010110001101101111\;, \\
K_{58}   & \deq & 10000100110010000111100000010100\;, \\
K_{59}   & \deq & 10001100110001110000001000001000\;, \\
K_{60}   & \deq & 10010000101111101111111111111010\;, \\
K_{61}   & \deq & 10100100010100000110110011101011\;, \\
K_{62}   & \deq & 10111110111110011010001111110111\;, \\
K_{63}   & \deq & 11000110011100010111100011110010\;.
\end{aeqns}
\end{ldispl}%

\end{document}